\documentclass[a4paper,12pt,eqno]{article}
\usepackage{theorem}
\usepackage{latexsym,amssymb,amsfonts,amsmath}
\usepackage{cite}

\newtheorem{thm}{Theorem}[section]

\newtheorem{pro}[thm]{Proposition}

\theoremheaderfont{\scshape}
\newcommand{\RM}{\mathbb{R}}

\newcommand{\PM}{\mathbb{P}}

\title{{\Large {\bf Continuous-time quantum walks on the threshold network model}}}
\author{
{\small Yusuke IDE}\\
{\scriptsize Department of Information Systems Creation,
Faculty of Engineering, 
Kanagawa University}\\
{\scriptsize Kanagawa, Yokohama 221-8686, Japan}\\
{\scriptsize Tel.: +81-45-481-5661, Fax: +81-45-413-6565}\\
{\scriptsize ide@kanagawa-u.ac.jp}\\
\\
{\small Norio KONNO}\\
{\scriptsize Department of Applied Mathematics, 
Faculty of Engineering, 
Yokohama National University}\\
{\scriptsize Hodogaya, Yokohama 240-8501, Japan}\\
{\scriptsize Tel.: +81-45-339-4205, Fax: +81-45-339-4205}\\
{\scriptsize konno@ynu.ac.jp}\\
}
\date{\empty }
\begin{document}
\maketitle

\par\noindent
\begin{small}
\par\noindent
{\bf Abstract}
It is well known that many real world networks have 
the power-law degree distribution (scale-free property).
However there are no rigorous results for continuous-time quantum walks on 
such realistic graphs. 
In this paper, we analyze  
space-time behaviors of continuous-time quantum walks and random walks 
on the threshold network model which is a reasonable candidate model 
having scale-free property. 
We show that the quantum walker exhibits localization at the starting point, 
although the random walker tends to spread uniformly.

\end{small}

\setcounter{equation}{0}
%%%%%%%%%%%%%%%%%%%%%%%%%%%%%%%%%%%%%%%%%%%%%%%%%%%%%%%%%%%%%%%%%%%%%%%%%%%%%%%%%%%%%%%%%%%
\section{Introduction}
Continuous-time quantum walks, which are the quantum counterparts of 
the classical random walks, 
have been widely studied on various deterministic graphs, such as 
the line \cite{Konno2005}, 
star graph \cite{Salimi2009AnnPhys,Xu2009}, 
cycle graph \cite{abtw2003,ikkk2005,MuelkenBlumen2006}, 
dendrimers \cite{mbb2006}, 
spidernet graphs \cite{Salimi2010QIP}, 
the Dual Sierpinski Gasket \cite{abm2008}, 
direct product of Cayley graphs \cite{SalimiJafarizadeh2009}, 
quotient graphs \cite{Salimi2008IJQI}, 
odd graphs \cite{Salimi2008IJTP}, 
trees \cite{Konno2006IDAQP,JafarizadehSalimi2007} 
and 
ultrametric spaces \cite{Konno2006IJQI}. 
For further information, see reviews such as \cite{Konno2008book,Venegas2008book}. 
Also there are simulation based study of continuous-time quantum walks 
on probabilistic graphs, such as
small-world networks \cite{mpb2007} and Erd\H{o}s-R\'enyi random graph \cite{XuLiu2008}. 
However there are no rigorous results for continuous-time quantum walks on 
such probabilistic graphs. 
In this paper, we focus on the continuous-time quantum walk on a random graph 
called the threshold network model. 

Many real world networks (graphs) are characterized by small diameters, high clustering, and 
power-law (scale-free) degree distributions \cite{Albert02,NewmanSIAM,blmch}. 
The threshold network model belongs to the so-called hidden variable models \cite{ccdm,so}
and is known for being capable of generating scale-free networks. 
Their mean behavior \cite{ccdm,so,bps,mmk04,scb,hss,fum} and limit theorems 
\cite{kmrs05,ikm07rims,ikm09MCAP,fikmmu09IIS} for the degree, the clustering coefficients, 
the number of subgraphs, and the average distance have been analyzed. 
The strong law of large numbers and central limit theorem for the rank of 
the adjacency matrix of the model with self-loops are given by \cite{BoseSen2007}. 
Eigenvalues and eigenvectors of the adjacency matrix \cite{iko10arXiv}, 
the Laplacian matrix \cite{Merris1994,Merris1998} of the model have been studied. 
See also \cite{mp95,kmrs05,mmk05,mk06,ikm07rims,dhj09,ikm09MCAP} for related works.

This paper is organized as follows. 
We define the threshold network model and give a brief review of the hierarchical structure 
of the graph in Section 2. In Section 3, we define the continuous-time quantum walk on the 
threshold network model and the special setting called the binary threshold model. The main 
results are presented in this section and the proofs are given in Section 4. 
Results on the continuous-time random walks on the models are obtained in Section 5. 
Summary is given in the last section.  
%%%%%%%%%%%%%%%%%%%%%%%%%%%%%%%%%%%%%%%%%%%%%%%%%%%%%%%%%%%%%%%%%%%%%%%%%%%%%%%%%%%%%%%%%%%
\section{Threshold network model}
The \textit{threshold network model} $\mathcal{G}_{n}(X,\theta )$ is a random graph 
on the vertex set $V=\{1,2,\ldots ,n\}$. 
Let $\{X_{1},X_{2},\ldots ,X_{n}\}$ be independent copies of a random variable $X$ 
with distribution $\PM$. We draw an edge between two distinct vertices $i,j\in V$ 
if $X_{i}+X_{j}>\theta $ where $\theta \in \RM$ is a constant called a threshold. 
Hereafter, we use $\PM^{\infty }$ as the distribution of $\{X_{i}\}_{i=1}^{\infty }$.

Each sample graph $G\in\mathcal{G}_{n}(X,\theta)$ has a hierarchical structure 
described by the so-called creation sequence \cite{dhj09,hss}.
Let 
$X_{(1)}\leq X_{(2)}\leq \cdots \leq X_{(n)}$
be a rearranged sequence of random variables 
$X_{1}, X_{2}, \ldots , X_{n}$ 
in increasing order. 
If $X_{(1)}+X_{(n)}>\theta$, we have
\[
\theta <X_{(1)}+X_{(n)}\leq X_{(2)}+X_{(n)}\leq 
\dots \leq X_{(n-1)}+X_{(n)},
\]
which means that the vertex corresponding to $X_{(n)}$ is connected with 
the $n-1$ other vertices. 
Otherwise, we have 
\begin{eqnarray*}
\theta\ge X_{(1)}+X_{(n)}\ge \dots \ge X_{(1)}+X_{(3)}
\ge X_{(1)}+X_{(2)},
\end{eqnarray*}
which means that the vertex corresponding to $X_{(1)}$ is isolatedD
We set $s_n=1$ or $s_n=0$ according as the former case or the latter occurs.
Then, according to the case we remove the random variable $X_{(n)}$ or $X_{(1)}$,
we continue similar procedure to define $s_{n-1},\dots,s_2$.
Finally, we set $s_1=s_2$ and obtain a $\{0,1\}$-sequence $\{s_1,s_2,\dots,s_n\}$,
which is called the \textit{creation sequence} of $G$ and is denoted by $S_G$.

Given a creation sequence $S_G$ let $k_{i}$ and $l_{i}$ denote the number of consecutive bits of $1$'s 
and $0$'s, respectively, as follows: 
\begin{equation}\label{dfkl}
S_{G}=\{
\overbrace{1,\ldots ,1}^{k_{1}},\overbrace{0,\ldots ,0}^{l_{1}},
\overbrace{1,\ldots ,1}^{k_{2}},\overbrace{0,\ldots ,0}^{l_{2}},
\ldots ,
\overbrace{1,\ldots ,1}^{k_{m}},\overbrace{0,\ldots ,0}^{l_{m}}
\}.
\end{equation}
It may happen that $k_{1}=0$ or $l_{m}=0$,
but we have $k_2,\dots,k_m,l_1,\dots,l_{m-1}\ge1$ and $m\ge1$.
Moreover, by definition we have two cases:
(a) $k_1=0$ (equivalently $s_1=0$) and $l_1\ge2$;
(b) $k_1\ge2$ (equivalently, $s_1=1$).

\begin{figure}
\begin{center}
%WinTpicVersion2.15
\unitlength 0.1in
\begin{picture}(16.00,14.00)(0.40,-14.00)
% DOT 0 0 3 0
% 9 400 805 400 1205 200 1605 600 1605 1200 805 1200 1205 1000 1605 1400 1605 1400 1605
% 
\special{pn 20}%
\special{sh 1}%
\special{ar 400 405 10 10 0  6.28318530717959E+0000}%
\special{sh 1}%
\special{ar 400 805 10 10 0  6.28318530717959E+0000}%
\special{sh 1}%
\special{ar 200 1205 10 10 0  6.28318530717959E+0000}%
\special{sh 1}%
\special{ar 600 1205 10 10 0  6.28318530717959E+0000}%
\special{sh 1}%
\special{ar 1200 405 10 10 0  6.28318530717959E+0000}%
\special{sh 1}%
\special{ar 1200 805 10 10 0  6.28318530717959E+0000}%
\special{sh 1}%
\special{ar 1000 1205 10 10 0  6.28318530717959E+0000}%
\special{sh 1}%
\special{ar 1400 1205 10 10 0  6.28318530717959E+0000}%
\special{sh 1}%
\special{ar 1400 1205 10 10 0  6.28318530717959E+0000}%
% STR 2 0 3 0
% 3 1200 400 1200 500 5 0
% $1$
\put(12.0000,-1.0000){\makebox(0,0){$1$}}%
% STR 2 0 3 0
% 3 400 410 400 510 5 0
% $0$
\put(4.0000,-1.1000){\makebox(0,0){$0$}}%
% LINE 2 2 3 0
% 16 40 600 1640 600 1640 1800 40 1800 40 1400 1640 1400 1640 1000 40 1000 40 400 40 1800 840 1800 840 400 1640 400 1640 1800 1640 400 40 400
% 
\special{pn 8}%
\special{pa 40 200}%
\special{pa 1640 200}%
\special{dt 0.045}%
\special{pa 1640 200}%
\special{pa 1639 200}%
\special{dt 0.045}%
\special{pa 1640 1400}%
\special{pa 40 1400}%
\special{dt 0.045}%
\special{pa 40 1400}%
\special{pa 41 1400}%
\special{dt 0.045}%
\special{pa 40 1000}%
\special{pa 1640 1000}%
\special{dt 0.045}%
\special{pa 1640 1000}%
\special{pa 1639 1000}%
\special{dt 0.045}%
\special{pa 1640 600}%
\special{pa 40 600}%
\special{dt 0.045}%
\special{pa 40 600}%
\special{pa 41 600}%
\special{dt 0.045}%
\special{pa 40 0}%
\special{pa 40 1400}%
\special{dt 0.045}%
\special{pa 40 1400}%
\special{pa 40 1399}%
\special{dt 0.045}%
\special{pa 840 1400}%
\special{pa 840 0}%
\special{dt 0.045}%
\special{pa 840 0}%
\special{pa 840 1}%
\special{dt 0.045}%
\special{pa 1640 0}%
\special{pa 1640 1400}%
\special{dt 0.045}%
\special{pa 1640 1400}%
\special{pa 1640 1399}%
\special{dt 0.045}%
\special{pa 1640 0}%
\special{pa 40 0}%
\special{dt 0.045}%
\special{pa 40 0}%
\special{pa 41 0}%
\special{dt 0.045}%
% LINE 2 0 3 0
% 22 1400 1610 1000 1610 1000 1610 1200 1210 1200 1210 1400 1610 1400 1610 1200 810 1200 810 1200 1210 1000 1610 1200 810 1200 810 400 1210 200 1610 1200 810 1200 810 600 1610 600 1610 1200 1210 1200 1210 200 1610
% 
\special{pn 8}%
\special{pa 1400 1210}%
\special{pa 1000 1210}%
\special{fp}%
\special{pa 1000 1210}%
\special{pa 1200 810}%
\special{fp}%
\special{pa 1200 810}%
\special{pa 1400 1210}%
\special{fp}%
\special{pa 1400 1210}%
\special{pa 1200 410}%
\special{fp}%
\special{pa 1200 410}%
\special{pa 1200 810}%
\special{fp}%
\special{pa 1000 1210}%
\special{pa 1200 410}%
\special{fp}%
\special{pa 1200 410}%
\special{pa 400 810}%
\special{fp}%
\special{pa 200 1210}%
\special{pa 1200 410}%
\special{fp}%
\special{pa 1200 410}%
\special{pa 600 1210}%
\special{fp}%
\special{pa 600 1210}%
\special{pa 1200 810}%
\special{fp}%
\special{pa 1200 810}%
\special{pa 200 1210}%
\special{fp}%
\end{picture}%
\caption{A threshold graph $G$ corresponding to $S_{G}=\{1,1,0,0,1,0,1,0\}$}
\label{kaisou3}
\end{center}
\end{figure}
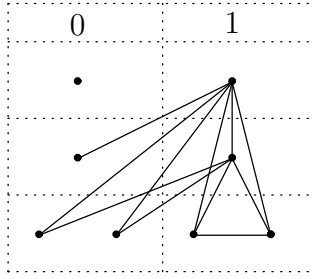

For example, if $S_{G}=\{1,1,0,0,1,0,1,0\}$ then 
$k_{1}=2,\ l_{1}=2,\ k_{2}=1,\ l_{2}=1,\ k_{3}=1,\ l_{3}=1$ and 
Fig.\ \ref{kaisou3} shows the shape of $G$. 
%--------------------------------------------------------------------------------------figure start
%--------------------------------------------------------------------------------------figure end

The creation sequence $S_G$ gives rise to a partition of the vertex set:
\[
V=\bigcup_{i=1}^m V^{(1)}_i \cup \bigcup_{i=1}^m V^{(0)}_i\,
\qquad |V^{(1)}_{i}|=k_i\,,
\quad |V^{(0)}_{i}|=l_i\,.
\]
The subgraph induced by $V^{(1)}_i$ is the complete graph of $k_i$ vertices,
and that induced by $V^{(0)}_i$ is the null graph of $l_{i}$ vertices.
Moreover, every vertex in $V^{(1)}_i$ (resp. $V^{(0)}_i$) is connected 
(resp. disconnected) with all vertices in 
\[
V^{(1)}_1\cup\dots\cup V^{(1)}_{i} \cup
V^{(0)}_1\cup\dots\cup V^{(0)}_{i-1}\,.
\]
In general, a graph possessing the above hierarchical structure is 
called a \textit{threshold graph} \cite{mp95}.
%%%%%%%%%%%%%%%%%%%%%%%%%%%%%%%%%%%%%%%%%%%%%%%%%%%%%%%%%%%%%%%%%%%%%%%%%%%%%%%%%%%%%%%%%%%
\section{Our model and results}
Let $A_{G}$ be the adjacency matrix and $D_{G}$ be the diagonal matrix of degrees 
(the sum of the rows of $A_{G}$) of $G\in \mathcal{G}_{n}(X,\theta )$. 
Then the Laplacian matrix $L_{G}$ of $G$ is given by $L_{G}=D_{G}-A_{G}$. 
The time evolution operator $U_{n,t}^{G}$ of a continuous-time quantum walk on 
$G$ is defined by 
\begin{eqnarray}\label{defU}
U_{n,t}^{G}=e^{itL_{G}}\equiv \sum _{k=0}^{\infty }\frac{(it)^{k}}{k!}L_{G}^{k}.
\end{eqnarray}

Let $\{\Psi _{n,t}^{G}\}_{t\geq 0}$ be the probability amplitude of the quantum walk, i.e.,
$
\Psi _{n,t}^{G}=U_{n,t}^{G}\Psi _{n,0}^{G},
$
and $X_{n,t}$ denotes the position of the quantum walker at time $t$. 
Then the probability that the quantum walker on $G$ is in position $x\in V$ 
at time $t$ with initial condition $\Psi _{n,0}^{G}$ is defined by 
\begin{eqnarray*}
P_{n,t}^{G}(X_{n,t}=x)\equiv |\Psi _{n,t}^{G}(x)|^{2},
\end{eqnarray*}
where 
$
\Psi _{n,t}^{G}=
{}^{T}\!
\begin{bmatrix}
\Psi _{n,t}^{G}(1) & \cdots & \Psi _{n,t}^{G}(n)
\end{bmatrix}
$. 
Here ${}^TA$ denotes the transpose of a matrix $A$. 

The time evolution operator is obtained as follows:
\begin{thm}\label{thmUtgeneral}
Suppose $G\in \mathcal{G}_{n}(X,\theta )$ is connected. 
The time evolution operator $U_{n,t}^{G}$ of the continuous-time quantum walk on 
$G$ is given by 
\begin{eqnarray*}\label{eqUtgeneral}
\left(U_{n,t}^{G}\right)_{v,w}=
\begin{cases}
\left(U_{n,t}^{G}\right)_{v,w}^{(1,i)}
& \text{if $v\in V_{i}^{(1)},\ w\in V_{j+1}^{(1)}\cup V_{j}^{(0)}\ (j\leq i-1)$},\\
\\
\left(U_{n,t}^{G}\right)_{v,w}^{(0,i)}
& \text{if $v\in V_{i}^{(0)}, w\in V_{j}^{(1)}\cup V_{j}^{(0)}\ (j\leq i)$}.
\end{cases}
\end{eqnarray*}
Here $A_{v,w}$ denotes the $(v,w)$ element of a matrix $A$ and 
\begin{align*}
\left(U_{n,t}^{G}\right)_{v,w}^{(1,i)}
&=
\left(
I_{V_{i}^{(1)}}(w)-\frac{1}{D_{k_{i}}-D_{l_{i}}+1}
\right)
e^{it(D_{k_{i}}+1)}
+
\displaystyle\sum_{j=i+1}^{m}
\frac{(D_{l_{j-1}}-D_{l_{j}})e^{it(D_{k_{j}}+1)}}
{(D_{k_{j}}-D_{l_{j-1}}+1)(D_{k_{j}}-D_{l_{j}}+1)}\\
&+
\displaystyle\sum_{j=i}^{m-1}
\frac{(D_{k_{j+1}}-D_{k_{j}})e^{itD_{l_{j}}}}
{(D_{k_{j}}-D_{l_{j}}+1)(D_{k_{j+1}}-D_{l_{j}}+1)}
+
\frac{1}{n},
\\
\left(U_{n,t}^{G}\right)_{v,w}^{(0,i)}
&=
\left(
I_{V_{i}^{(0)}}(w)-\frac{1}{D_{k_{i+1}}-D_{l_{i}}+1}
\right)
e^{itD_{l_{i}}}
+
\displaystyle\sum_{j=i}^{m}
\frac{(D_{l_{j-1}}-D_{l_{j}})e^{it(D_{k_{j}}+1)}}
{(D_{k_{j}}-D_{l_{j-1}}+1)(D_{k_{j}}-D_{l_{j}}+1)}\\
&+
\displaystyle\sum_{j=i+1}^{m-1}
\frac{(D_{k_{j+1}}-D_{k_{j}})e^{itD_{l_{j}}}}
{(D_{k_{j}}-D_{l_{j}}+1)(D_{k_{j+1}}-D_{l_{j}}+1)}
+\frac{1}{n}. 
\end{align*}
Where $D_{k_{i}}$ and $D_{l_{i}}$ denote the degree of vertices in $V_{i}^{(1)}$ 
and $V_{i}^{(0)}$, respectively and $I_{A}(x)$ is the indicator function of a set $A$, i.e., 
$I_{A}(x)=1$ if $x\in A$ and $I_{A}(x)=0$ otherwise. 
\end{thm}

Theorem \ref{thmUtgeneral} shows that we can obtain the probability of the quantum walker 
in position $x\in V$ at time $t$ for any initial conditions at least in principle. 
But in general cases, it is hard to obtain the probability. 
In this paper, 
we analyze behaviors of quantum walks starting from a vertex 
$v$, i.e., the case of 
\begin{eqnarray*}
\Psi_{n,0}(s)=
\begin{cases}
1, & \text{if }s=v,\\
0, & \text{otherwise}.
\end{cases}
\end{eqnarray*} 

\begin{thm}\label{thmPtgeneral}
Suppose $G$ is connected. 
The limit of the probability of the quantum walker starting from a vertex 
$v\in V_{m}^{(1)}$ is given by 
\begin{eqnarray*}\label{limPtgeneral}
\lim_{n\to \infty }P_{n,t}^{G}(x)=
\begin{cases}
1, & \text{if }x=v,\\
0, & \text{otherwise,}
\end{cases}
\quad \text{for $\PM^{\infty }$-almost every $G$.}
\end{eqnarray*}
\end{thm}

Because $P_{n,t}^{G}$ is not converge 
in $t\to \infty $, 
we study the time-averaged probability $\bar{P}_{n}^{G}(x)$ defined by 
\begin{eqnarray*}
\bar{P}_{n}^{G}(x)=
\lim _{T\to \infty }\frac{1}{T}\int _{0}^{T}P_{n,t}^{G}(X_{n,t}=x)dt.
\end{eqnarray*}
\begin{thm}\label{thmPtbargeneral}
Suppose $G\in \mathcal{G}_{n}(X,\theta )$ is connected. 
The time-averaged probability $\bar{P}_{n}^{G}(x)$ 
of the quantum walker on $G$ starting from a vertex $v\in V_{m}^{(1)}$
is 
\begin{eqnarray*}
\bar{P}_{n}^{G}(x)=
\begin{cases}
\left(1-\frac{1}{n}\right)^{2}+\frac{1}{n^{2}}, & \text{if }x=v,\\
\frac{2}{n^{2}}, & \text{otherwise}.
\end{cases}
\end{eqnarray*}
\end{thm}

In order to study more detailed properties of the quantum walk on the model, 
we focus on the threshold network model $\mathcal{G}_{n}(X,\theta )$ 
defined by Bernoulli trials with success probability $p\in (0,1)$, i.e., 
$\PM(X=1)=1-\PM(X=0)=p$, and a threshold $\theta \in [0,1)$. 
We call this model the \textit{binary threshold model} $\mathcal{G}_{n}(p)$. 
For each $G\in \mathcal{G}_{n}(p)$, i.e., realization $G$ of $\mathcal{G}_{n}(p)$, 
we consider a partition of the vertex set $V$:
\begin{eqnarray*}
V=V_{G}^{(1)}\cup V_{G}^{(0)},\quad 
V_{G}^{(1)}=\{i:X_{i}=1\},\quad 
V_{G}^{(0)}=\{i:X_{i}=0\}.
\end{eqnarray*}
It is easy to see that the subgraph induced by $V_{G}^{(1)}$ is the complete graph 
on $k_{G}\equiv \sharp V_{G}^{(1)}$ vertices, and that induced by $V_{G}^{(0)}$ is
the null graph on $l_{G}\equiv \sharp V_{G}^{(0)}$ vertices, where $\sharp A$ is 
the number of elements in a set $A$. Moreover, every vertex 
in $V_{G}^{(1)}$ is connected with all vertices in $V_{G}^{(0)}$. 
Note that this is the case of $m=2$, $k_{1}=l_{2}=0$, $k_{2}=k_{G}$ and $l_{1}=l_{G}$ 
in Eq.\ (\ref{dfkl}). 

The time evolution operator of the continuous-time quantum walk on the binary threshold model 
is obtained as follows:
\begin{thm}\label{thmUt}
The time evolution operator $U_{n,t}^{G}$ of the continuous-time quantum walk on 
$G\in \mathcal{G}_{n}(p)$ is given by 
\begin{eqnarray*}\label{eqUt}
U_{n,t}^{G}=
\begin{bmatrix}
U_{k_{G},k_{G}} & U_{k_{G},l_{G}}\\
U_{l_{G},k_{G}} & U_{l_{G},l_{G}}
\end{bmatrix}. 
\end{eqnarray*}
The elements of $U_{n,t}^{G}$ are 
\begin{align*}
U_{k_{G},k_{G}}
&=
e^{int}I_{k_{G}}+\frac{1-e^{int}}{n}\mathbf{1}_{k_{G},k_{G}},\\
U_{k_{G},l_{G}}
&=
\frac{1-e^{int}}{n}\mathbf{1}_{k_{G},l_{G}},\\
U_{l_{G},k_{G}}
&=
\frac{1-e^{int}}{n}\mathbf{1}_{l_{G},k_{G}},\\
U_{l_{G},l_{G}}
&=
e^{ik_{G}t}I_{l_{G}}+
\left(\frac{1}{n}+\frac{k_{G}e^{int}}{nl_{G}}-\frac{e^{ik_{G}t}}{l_{G}}\right)\mathbf{1}_{l_{G},l_{G}},
\end{align*}
where $\mathbf{1}_{i,j}$ is the $i\times j$ matrix consisting of only $1$ 
and $I_{i}$ is the $i\times i$ identity matrix. 
\end{thm}

On the binary threshold model, a strong localization is observed at any starting point as follows:  
\begin{pro}\label{thmPt}
The limit of the probability of the quantum walker starting from a vertex $v\in V$ is given by 
\begin{eqnarray*}\label{limPt}
\lim_{n\to \infty }P_{n,t}^{G}(x)=
\begin{cases}
1, & \text{if }x=v,\\
0, & \text{otherwise,}
\end{cases}
\quad \text{for $\PM^{\infty }$-almost every $G$.}
\end{eqnarray*}
\end{pro}
Note that if the quantum walker starts from a vertex $v\in V_{G}^{(1)}$ then 
the statement of Proposition \ref{thmPt} is the same as Theorem \ref{thmPtgeneral}. 

The time averaged probability of the quantum walker on the binary threshold model 
is obtained as follows: 
\begin{pro}\label{thmPtbar}
The time-averaged probability $\bar{P}_{n}^{G}(x)$ 
of the quantum walker on $G\in \mathcal{G}_{n}(p)$
starting from a vertex $v\in V_{G}^{(0)}$ is given by 
\begin{eqnarray*}
\bar{P}_{n}^{G}(x)=
\begin{cases}
\left(1-\frac{1}{l_{G}}\right)^{2}
+\left(\frac{k_{G}}{nl_{G}}\right)^{2}+\frac{1}{n^{2}}, & \text{if }x=v,\\
\frac{1}{l_{G}^{2}}
+\left(\frac{k_{G}}{nl_{G}}\right)^{2}+\frac{1}{n^{2}}, & \text{if }x\in V_{G}^{(0)}\setminus \{v\},\\
\frac{2}{n^{2}}, & \text{otherwise}.
\end{cases}
\end{eqnarray*}
\end{pro}
Note that when the quantum walker starts from a vertex $v\in V_{G}^{(1)}$, we can use 
Theorem \ref{thmPtbargeneral}. 
%%%%%%%%%%%%%%%%%%%%%%%%%%%%%%%%%%%%%%%%%%%%%%%%%%%%%%%%%%%%%%%%%%%%%%%%%%%%%%%%%%%%%%%%%%%
\section{Proofs}
The Laplacian matrix $L_{G}$ of $G\in \mathcal{G}_{n}(p)$ 
is given by 
\begin{eqnarray*}
L_{G}=
\begin{bmatrix}
nI_{k_{G}}-\mathbf{1}_{k_{G},k_{G}} & -\mathbf{1}_{k_{G},l_{G}}\\
-\mathbf{1}_{l_{G},k_{G}} & kI_{l_{G}}
\end{bmatrix}.
\end{eqnarray*}
Eigenvalues and eigenvectors of $L_{G}$ are known as follows \cite{hss,Merris1994,Merris1998}:
\begin{center}
\begin{tabular}[l]{|l|l|}
\hline
eigenvalue & eigenvectors\\
\hline
$n$
& 
$
\begin{array}{l}
\\ 
\mathbf{v}_{j}\equiv 
\frac{1}{\sqrt{j(j+1)}}
\begin{bmatrix}
\mathbf{1}_{j,1} \\ -j \\ \mathbf{0}_{n-j-1,1}
\end{bmatrix}
\ (1\leq j\leq k_{G}-1), 
\mathbf{v}_{k_{G}}\equiv 
\frac{1}{\sqrt{nk_{G}l_{G}}}
\begin{bmatrix}
l_{G}\mathbf{1}_{k_{G},1} \\ -k_{G}\mathbf{1}_{l_{G},1}
\end{bmatrix},\\
\quad 
\end{array}
$
\\
\hline 
$k_{G}$
& 
$
\begin{array}{l}
\\ 
\mathbf{w}_{j}\equiv 
\frac{1}{\sqrt{j(j+1)}}
\begin{bmatrix}
\mathbf{0}_{k_{G},1} \\ \mathbf{1}_{j,1} \\ -j \\ \mathbf{0}_{l_{G}-j-1,1}
\end{bmatrix}
\ (1\leq j\leq l_{G}-1),\\
\quad 
\end{array}
$
\\
\hline
$0$
& 
$
\begin{array}{l}
\\ 
\mathbf{w}_{l_{G}}\equiv
\frac{1}{\sqrt{n}}
\begin{bmatrix}
\mathbf{1}_{n,1}
\end{bmatrix},\\
\quad 
\end{array}
$
\\
\hline
\end{tabular}
\end{center}
where $\mathbf{0}_{i,j}$ is the $i\times j$ zero matrix. 
Note that the set of these eigenvectors forms an orthonormal basis of $\RM^{n}$. 
Thus we can define a orthogonal matrix $B_{G}$ corresponding to the eigenvectors as follows:
\begin{eqnarray}\label{defB}
B_{G}=
\begin{bmatrix}
\mathbf{v}_{1} & \ldots & \mathbf{v}_{k_{G}} & \mathbf{w}_{1} & \ldots & \mathbf{w}_{l_{G}}
\end{bmatrix}.
\end{eqnarray}

Because $U_{n,t}^{G}$ is an $n\times n$ (finite) matrix, using Eqs.\ (\ref{defU}) and (\ref{defB}), 
it is represented by 
\begin{eqnarray*}
U_{n,t}^{G}=B_{G}
\begin{bmatrix}
e^{int}I_{k_{G}} & \mathbf{0}_{k_{G},l_{G}} & \mathbf{0}_{k_{G},1}\\
\mathbf{0}_{l_{G}-1,k_{G}} & e^{ik_{G}t}I_{l_{G}-1} & \mathbf{0}_{l_{G}-1,1}\\
\mathbf{0}_{1,k_{G}} & \mathbf{0}_{1,l_{G}-1} & 1
\end{bmatrix}
{}^T\!B_{G}.
\end{eqnarray*}
It is easy to see that
\begin{eqnarray*}
U_{n,t}^{G}=
\begin{bmatrix}
e^{int}I_{k_{G}}
+\left[\left(-\frac{1}{k_{G}}+\frac{l_{G}}{nk_{G}}\right)e^{int}
+\frac{1}{n}\right]\mathbf{1}_{k_{G},k_{G}} & \frac{1}{n}\left(1-e^{int}\right)\mathbf{1}_{k_{G},l_{G}}\\
\frac{1}{n}\left(1-e^{int}\right)\mathbf{1}_{l_{G},k_{G}} & 
e^{ik_{G}t}I_{k_{G}}
+\left[-\frac{1}{l_{G}}e^{ik_{G}t}+\frac{k_{G}}{nl_{G}}e^{int}
+\frac{1}{n}\right]\mathbf{1}_{l_{G},l_{G}}
\end{bmatrix}.
\end{eqnarray*}
By using a relation
$-1/k_{G}+l_{G}/nk_{G}=-1/n$,
we obtain Theorem \ref{thmUt}. 

It is given by simple calculations that 
\begin{align*}
\left|\left(1-\frac{1}{n}\right)e^{int}+\frac{1}{n}\right|^{2}
&=1-\frac{2}{n}\left(1-\frac{1}{n}\right)\left(1-\cos nt\right),\\
\left|\frac{1}{n}\left(1-e^{int}\right)\right|^{2}
&=\frac{1}{n^{2}}\left(2-\cos nt\right),\\
\left|\left(1-\frac{1}{l_{G}}\right)e^{ik_{G}t}+\frac{k_{G}e^{int}}{nl_{G}}+\frac{1}{n}\right|^{2}
&=\left(1-\frac{1}{l_{G}}\right)^{2}+\left(\frac{k_{G}}{nl_{G}}\right)^{2}+\frac{1}{n^{2}}\\
&+\frac{2k_{G}}{nl_{G}}\left(1-\frac{1}{l_{G}}\right)\left(\cos l_{G}t+\frac{l_{G}}{k_{G}}\cos k_{G}t\right)
+\frac{2k_{G}}{n^{2}l_{G}}\cos nt,\\
\left|-\frac{e^{ik_{G}t}}{l_{G}}+\frac{k_{G}e^{int}}{nl_{G}}+\frac{1}{n}\right|^{2}
&=\frac{1}{l_{G}^{2}}+\left(\frac{k_{G}}{nl_{G}}\right)^{2}+\frac{1}{n^{2}}\\
&-\frac{2k_{G}}{nl_{G}^{2}}
\left(\cos l_{G}t+\frac{l_{G}}{k_{G}}\cos k_{G}t-\frac{l_{G}}{n}\cos nt
\right).
\end{align*}
On the other hand, we see that $\lim _{n\to \infty }k_{G}/n=p$ and $\lim _{n\to \infty }l_{G}/n=1-p$, 
$\PM^{\infty}$-almost surely 
by the strong law of large numbers for the i.i.d.\ Bernoulli sequence. 
Combining these facts and Theorem \ref{thmUt}, we have Proposition \ref{thmPt}. 
Also we obtain Proposition \ref{thmPtbar} immediately from
\begin{eqnarray*}
\lim _{T\to \infty }\frac{1}{T}\int _{0}^{T}\cos nt\ dt=
\lim _{T\to \infty }\frac{1}{T}\int _{0}^{T}\cos k_{G}t\ dt=
\lim _{T\to \infty }\frac{1}{T}\int _{0}^{T}\cos l_{G}t\ dt=
0.
\end{eqnarray*}

In the case of $G\in \mathcal{G}_{n}(X,\theta )$ which is connected, 
eigenvalues and eigenvectors of $L_{G}$ are also known as follows \cite{hss,Merris1994,Merris1998}:
\begin{center}
\begin{tabular}[l]{|l|l|}
\hline
eigenvalue & eigenvectors\\
\hline
$
\begin{array}{l}
D_{k_{i}}+1\\
(2\leq i\leq m)
\end{array}
$
& 
$
\begin{array}{l}
\\ 
\frac{1}{\sqrt{j(j+1)}}
\begin{bmatrix}
\mathbf{0}_{u_{i}+l_{i}} \\ \mathbf{1}_{j,1} \\ -j \\ \mathbf{0}_{d_{i}+k_{i}-j-1,1}
\end{bmatrix}
\ (1\leq j\leq k_{i}-1),\\
\frac{1}{\sqrt{(k_{i}+d_{i})k_{i}d_{i}}}
\begin{bmatrix}
\mathbf{0}_{u_{i}+l_{i}} \\ d_{i}\mathbf{1}_{k_{i},1} \\ -k_{i}\mathbf{1}_{d_{i},1}
\end{bmatrix},\\
\quad 
\end{array}
$
\\
\hline 
$
\begin{array}{l}
D_{l_{i}}\\
(2\leq i\leq m-1)
\end{array}
$
& 
$
\begin{array}{l}
\\ 
\frac{1}{\sqrt{j(j+1)}}
\begin{bmatrix}
\mathbf{0}_{u_{i},1} \\ \mathbf{1}_{j,1} \\ -j \\ \mathbf{0}_{d_{i+1}-j-1,1}
\end{bmatrix}
\ (1\leq j\leq l_{i}-1),\\
\frac{1}{\sqrt{(k_{i}+d_{i})l_{i}d_{i+1}}}
\begin{bmatrix}
\mathbf{0}_{u_{i}+l_{i}} \\ (k_{i}+d_{i})\mathbf{1}_{l_{i},1} \\ -l_{i}\mathbf{1}_{k_{i}+d_{i},1}
\end{bmatrix},\\
\quad 
\end{array}
$
\\
\hline
$D_{k_{1}}+1$
& 
$
\begin{array}{l}
\\ 
\frac{1}{\sqrt{j(j+1)}}
\begin{bmatrix}
\mathbf{0}_{u_{i}+l_{i}} \\ \mathbf{1}_{j,1} \\ -j \\ \mathbf{0}_{k_{1}-j-1,1}
\end{bmatrix}
\ (j=1,\ldots ,k_{1}-1),\\
\quad 
\end{array}
$
if $k_{1}\neq 0$, 
\\
\hline 
$D_{l_{1}}$
& 
$
\begin{array}{l}
\\ 
\frac{1}{\sqrt{j(j+1)}}
\begin{bmatrix}
\mathbf{0}_{u_{1},1} \\ \mathbf{1}_{j,1} \\ -j \\ \mathbf{0}_{d_{2}-j-1,1}
\end{bmatrix}
\ (j=1,\ldots ,l_{1}-1),\\
\quad 
\end{array}
$
\\
\hline 
$D_{l_{1}}$
& 
$
\begin{array}{l}
\\ 
\frac{1}{\sqrt{(k_{1}+d_{1})l_{1}d_{2}}}
\begin{bmatrix}
\mathbf{0}_{u_{1}+l_{1}} \\ (k_{1}+d_{1})\mathbf{1}_{l_{1},1} \\ -l_{1}\mathbf{1}_{k_{1}+d_{1},1}
\end{bmatrix},\\
\quad 
\end{array}
$
if $k_{1}\neq 0$, 
\\
\hline
$0$
& 
$
\begin{array}{l}
\\ 
\frac{1}{\sqrt{n}}
\begin{bmatrix}
\mathbf{1}_{n,1}
\end{bmatrix},\\
\quad 
\end{array}
$
\\
\hline
\end{tabular}
\end{center}
where $u_{i}=\sum _{j>i}(k_{j}+l_{j})$ and $d_{i}=\sum _{j<i}(k_{j}+l_{j})$. 
By the same argument in the proof of Theorem \ref{thmUt} and the following relations:
\begin{align*}
k_{1}& =D_{k_{1}}-D_{l_{1}}+1\ (\text{if}\ k_{1}\neq 0),\\
k_{i}& =D_{l_{i-1}}-D_{l_{i}}\ (2\leq i\leq m),\\
l_{i}& =D_{k_{i+1}}-D_{k_{i}}\ (1\leq i\leq m-1),
\end{align*}
we have Theorem \ref{thmUtgeneral}. 

Note that $D_{k_{m}}=n-1$ and $D_{l_{m}}=0$ by assumpution. 
Comparing Theorem \ref{thmUtgeneral} with Theorem \ref{thmUt}, 
the transition probability of the walker on $G\in \mathcal{G}_{n}(p)$ starting from a vertex 
$v\in V_{m}^{(1)}$ is the same as that of on $G\in \mathcal{G}_{n}(p)$ starting from 
a vertex $v\in V_{G}^{(1)}$. Thus we have Theorems \ref{thmPtgeneral} and \ref{thmPtbargeneral}.
%%%%%%%%%%%%%%%%%%%%%%%%%%%%%%%%%%%%%%%%%%%%%%%%%%%%%%%%%%%%%%%%%%%%%%%%%%%%%%%%%%%%%%%%%%%
\section{Classical Case}
The time evolution operator $\mathcal{U}_{n,t}^{G}$ of a continuous-time random walk on 
$G\in \mathcal{G}_{n}(X,\theta )$ is defined by 
\begin{eqnarray*}
\mathcal{U}_{n,t}^{G}=e^{-tL_{G}}\equiv \sum _{k=0}^{\infty }\frac{(-t)^{k}}{k!}L_{G}^{k}. 
\end{eqnarray*}
Let $\{\mathcal{P}_{n,t}^{G}\}_{t\geq 0}$ be the probability distribution of the random walk, i.e.,
$
\mathcal{P}_{n,t}^{G}=\mathcal{U}_{n,t}^{G}\mathcal{P}_{n,0}^{G},
$
and 
$Y_{n,t}$ denotes the position of the random walker at time $t$. 
By the same observation in the previous section, we have the same results 
for $\mathcal{U}_{n,t}^{G}$ as Theorems \ref{thmUtgeneral} and \ref{thmUt} 
by exchanging $it$ of $U_{n,t}^{G}$ for $t$. 
Using these results, we have the following:
\begin{pro}\label{thmCPt}
The limit of the probability that the random walker starting from a vertex $v\in V_{m}^{(1)}$ 
is given by 
\begin{eqnarray*}\label{limCPt}
\lim_{n\to \infty }n\mathcal{P}_{n,t}^{G}(y)=
1,\quad \text{for all $y\in V$, for $\PM^{\infty }$-almost every $G$.}
\end{eqnarray*}
\end{pro}
\begin{pro}\label{thmCPtbar}
The long-time limit of the probability of the random walk on 
$G\in \mathcal{G}_{n}(p)$ starting from a vertex $v\in V$ 
is given by 
\begin{eqnarray*}
\lim_{t\to \infty }\mathcal{P}_{n,t}^{G}(y)=\frac{1}{n}
,\quad \text{for all $y\in V$}.
\end{eqnarray*}
\end{pro}
We can also estimate the time-averaged probability $\bar{\mathcal{P}}_{n}^{G}(y)$.
By simple calculation, we have $\bar{\mathcal{P}}_{n}^{G}(y)=1/n$ for a random walk on 
$G\in \mathcal{G}_{n}(p)$ starting from a vertex $v\in V$.
%%%%%%%%%%%%%%%%%%%%%%%%%%%%%%%%%%%%%%%%%%%%%%%%%%%%%%%%%%%%%%%%%%%%%%%%%%%%%%%%%%%%%%%%%%%
\section{Summary}
In this paper, we study the continuous-time quantum and random walks 
on the threshold network model. 
By comparing Theorem \ref{thmPtgeneral} with Proposition \ref{thmCPt}, we have quite different 
limit behaviors in $n\to \infty $ for the two types of walks starting from a vertex which degree equals $n-1$. 
Although quantum walkers exhibit strong localization at 
the starting point, random walkers tend to spread uniformly. 

Theorem \ref{thmPtbargeneral} and Proposition \ref{thmPtbar} show that the time-averaged probabilities 
of quantum walkers are not 
the uniform distribution (different from random walks). Furthermore, the time-averaged probability 
shows localization at starting point as $n\to \infty $. 
In the case of the binary threshold model, 
the rate of convergence are slightly 
different in the two starting points. Indeed, we obtain 
$
\lim _{n\to \infty }n(1-\bar{P}_{n}^{G}(v_{1}))=2 < 
\lim _{n\to \infty }n(1-\bar{P}_{n}^{G}(v_{0}))=2/(1-p)
$, $\PM^{\infty }$-almost surely for 
$v_{0}\in V_{G}^{(0)}$ and $v_{1}\in V_{G}^{(1)}$. 
A study covering the more general setting is now in progress.
%%%%%%%%%%%%%%%%%%%%%%%%%%%%%%%%%%%%%%%%%%%%%%%%%%%%%%%%%%%%%%%%%%%%%%%%%%%%%%%%%%%%%%%%%%%
\\
\\
{\bf Acknowledgment.} 
NK is funded by 
the Grant-in-Aid for Scientific Research (C) of 
Japan Society for the Promotion of Science (Grant No. 21540118). 
%%%%%%%%%%%%%%%%%%%%%%%%%%%%%%%%%%%%%%%%%%%%%%%%%%%%%%%%%%%%%%%%%%%%%%%%%%%%%%%%%%%%%%%%%%%

%%%%%%%%%%%%%%%%%%%%%%%%%%%%%%%%%%%%%%%%%%%%%%%%%%%%%%%%%%%%%%%%%%%%%%%%%%%%%%%%%%%%%%%%%%%%%
\end{document}